%% file: main.tex
\documentclass[conference]{IEEEtran}
\IEEEoverridecommandlockouts
\usepackage{cite}
\usepackage{bm}
\usepackage{amsmath,amssymb,amsfonts}
\usepackage{algorithmic}
\usepackage{graphicx}
\usepackage{textcomp}
\usepackage{enumitem}
\usepackage{dblfloatfix}
\usepackage{fixltx2e}
\usepackage[dvipsnames]{xcolor}
\usepackage{siunitx}
\usepackage{subcaption}
\usepackage{siunitx}

\usepackage[top=0.7in, bottom=1.0in, left=0.625in, right=0.625in]{geometry} % ICC submission edas
\usepackage[T1]{fontenc}
\usepackage{newtxtext,newtxmath}
\usepackage[hidelinks,bookmarks=false]{hyperref}

\usepackage{float}% better control of floating figures and tables
\usepackage{adjustbox}
\usepackage{booktabs}
\usepackage{tabularx}
\usepackage{makecell}
\usepackage{tikz}
\usepackage{graphicx}
\usepackage{caption}
\usepackage{hyperref}
\usepackage{array}
\usepackage{multirow}
\newcommand{\wh}[1]{\widehat{#1}}

\newcommand{\bsym}{\boldsymbol}
\newcolumntype{K}{>{\raggedright\arraybackslash}m{2.2 cm}}
\newcolumntype{E}{>{\centering\arraybackslash}m{2.3 cm}}
\newcolumntype{Y}{>{\centering\arraybackslash}m{4.2 cm}}
\newcolumntype{C}{>{\centering\arraybackslash}m{3.1 cm}}
\newcolumntype{d}{>{\centering\arraybackslash}m{1.35 cm}}
\newcolumntype{a}{>{\centering\arraybackslash}m{1.2 cm}}
\newcolumntype{e}{>{\centering\arraybackslash}m{1.35 cm}}

\newcolumntype{D}{>{\centering\arraybackslash}m{3 cm}}
\newcolumntype{z}{>{\centering\arraybackslash}m{0.7 cm}}
\newcolumntype{L}{>{\centering\arraybackslash}m{1 cm}}

\newcolumntype{x}{>{\centering\arraybackslash}m{0.75 cm}}
\newcolumntype{P}{>{\centering\arraybackslash}m{1.5 cm}}

\def\BibTeX{{\rm B\kern-.05em{\sc i\kern-.025em b}\kern-.08em
    T\kern-.1667em\lower.7ex\hbox{E}\kern-.125emX}}

\newcommand{\subsf}{\sf \scriptscriptstyle}
\newcommand{\mc}{\mathcal}
\def\SNR    {{\mathsf{SNR}}}
\def\INR    {{\mathsf{INR}}}

\newcommand{\RX}[1]{\mathrm{RX}_{#1}}

\begin{document}

\title{Joint Detection, Channel Estimation and Interference Nulling for Terrestrial-Satellite Downlink Co-Existence in the Upper Mid-Band}
%Upper Mid-Band Interference Nulling

    \author{
    \IEEEauthorblockN{Shizhen Jia\IEEEauthorrefmark{1}, Mingjun Ying\IEEEauthorrefmark{1},  Marco Mezzavilla\IEEEauthorrefmark{2}, Doru Calin\IEEEauthorrefmark{3}, Theodore S. Rappaport\IEEEauthorrefmark{1}, and Sundeep Rangan\IEEEauthorrefmark{1}}

    \IEEEauthorblockA{\IEEEauthorrefmark{1}NYU WIRELESS, New York University, Brooklyn, NY, USA 
    }
    \IEEEauthorblockA{\IEEEauthorrefmark{2}Dipartimento di Elettronica, Informazione e Bioingegneria (DEIB), Politecnico di Milano, Milan, Italy 
    % \{marco.mezzavilla\}@polimi.it
        }
    \IEEEauthorblockA{\IEEEauthorrefmark{3}MediaTek USA Inc., Warren, NJ, USA \\
    \{s.jia, srangan\}@nyu.edu
    % \{doru.calin\}@mediatek.com
    }
    \thanks{This work was supported, in part, by NSF grants 2345139, 2148293, 2133662, 1952180, 1904648, the NTIA Public Wireless Innovation Fund, and the industrial affiliates of NYU WIRELESS. Code is available at https://github.com/SJ00425/NTN-Nulling}
}

\maketitle
\begin{abstract}
The upper mid-band FR3 spectrum (7--24\,GHz) has garnered significant interest for future cellular services. However, utilizing a large portion of this band requires careful interference coordination with incumbent satellite systems. This paper investigates interference from high-power terrestrial base stations (TN-BSs) to satellite downlink receivers. A central challenge is that the victim receivers, i.e., ground-based non-terrestrial user equipment (NTN-UEs), such as satellite customer premises equipment, must first be detected, and their channels estimated, before the TN-BS can effectively place nulls in their directions. We explore a potential solution where NTN-UEs periodically transmit preambles or beacon signals that TN-BSs can use for detection and channel estimatio. The performance of this nulling approach is analyzed in a simplified scenario with a single victim, revealing the interplay between path loss and estimation quality in determining nulling performance. To further validate the method, we conduct a detailed multi-user site-specific ray-tracing (RT) simulation in a rural environment. The results show that the proposed nulling approach is effective under realistic parameters, even with high densities of victim units, although TN-BS may require a substantial number of antennas.

% The upper mid-band FR3 spectrum (7--24\,GHz) has garnered significant interest for future cellular services. However, utilizing a large portion of this band requires careful within-band interference coordination with incumbent satellite systems. This paper investigates interference from high-power terrestrial base stations (TN-BSs) to satellite downlink receivers. A central challenge is that the victim receivers, i.e., ground-based non-terrestrial user equipment (NTN-UEs), such as satellite customer premises equipment, must first be detected, and their channels estimated, before the TN-BS can effectively place nulls in their directions. We explore a potential solution where NTN-UEs periodically transmit preambles or beacon signals that TN-BSs can use for detection and channel estimation. The performance of the interference nulling approach is analyzed in a simplified scenario with a single victim, providing information on the effects of path loss and estimation overhead. To further validate the method, we conduct a detailed multi-user site-specific ray-tracing (RT) simulation in a rural environment. The RT simulation results show that the proposed nulling approach is effective under realistic parameters, even with high densities of victim units, although TN-BS may require a substantial number of antennas.

\end{abstract}


\input{Introduction}\label{sec:intro}

\input{Problem}\label{sec:prob}

\input{results}\label{results}

\section{Conclusion}
\label{conclusion}

In this paper, we proposed an interference nulling framework that effectively addresses TN-BSs' downlink interference towards NTN-UEs using a four-step protocol involving beacon broadcast, victim detection, channel estimation, and interference nulling. Simulations revealed a key trade-off between interference power and detection quality in determining nulling performance. 
Detailed Sionna RT evaluations confirmed the practicality of the approach, demonstrating substantial interference reduction, with future work exploring calibration-based refinements to the RT process to better account for real-world propagation effects and improve accuracy. Our results also underline the importance of large-scale antenna arrays (64+ elements) at TN-BS to effectively manage high densities of NTN-UEs, ensuring reliable and efficient spectrum sharing for future 5G and 6G systems.

% \medskip
% \emph{Acknowledgements}:  This work was supported, in part, by
% NSF grants 2345139, 2148293, 2133662, 1952180, 1904648,
% the NTIA Public Wireless Innovation Fund, and the industrial affiliates
% of NYU Wireless.

\bibliographystyle{IEEEtran}
\bibliography{references}

\end{document}

%% file: Introduction.tex
\section{Introduction}

The FR3 upper mid-band spectrum (7--\SI{24}{GHz}) offers unique value for cellular networks by balancing coverage and capacity needs in 5G and 6G technologies \cite{kang2024cellular, Shakya2024ojcom}, and has been identified by the 2023 World Radiocommunication Conference (WRC-23) as an essential band for future expansion \cite{NTIA2024}. A key challenge arises as FR3 overlaps with the satellite Ku-band (12--\SI{18}{GHz}), where commercial satellite systems operate as incumbents. The Federal Communications Commission (FCC) has proposed spectrum sharing rules to enable coexistence between terrestrial networks (TN) and non-terrestrial networks (NTN) via interference mitigation protocols \cite{FCC2024}. Recent theoretical work further shows such coexistence is feasible under clustered TN deployments using tractable outage models \cite{2024tractable}.

% The FR3 upper mid-band spectrum (7--\SI{24}{GHz}) provides unique value for cellular networks by balancing coverage and capacity needs for 5G and 6G technologies \cite{kang2024cellular,Shakya2024ojcom,Ying2025tcom}, and has been identified by the World Radiocommunication Conference 2023 (WRC-23)  as essential band for future cellular expansion \cite{NTIA2024}. A critical challenge emerges as FR3 overlaps with satellite Ku-band (12--\SI{18}{GHz}) where commercial satellite systems operate as incumbents. The Federal Communications Commission (FCC) proposed spectrum sharing regulations to enable terrestrial networks (TN) and non-terrestrial networks (NTN) coexistence through interference mitigation protocols \cite{FCC2024}. Recent theoretical analysis has further shown that such coexistence is feasible under clustered TN deployments using tractable outage models \cite{2024tractable}.

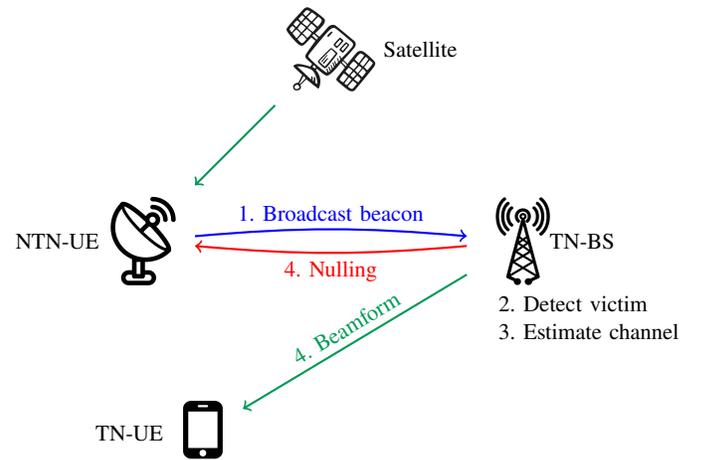
\begin{figure}[t]
    \centering
    \begin{tikzpicture}[scale=0.85, transform shape]
        % Satellite
        \node (satellite) at (4,6) {\includegraphics[width=1.5cm]{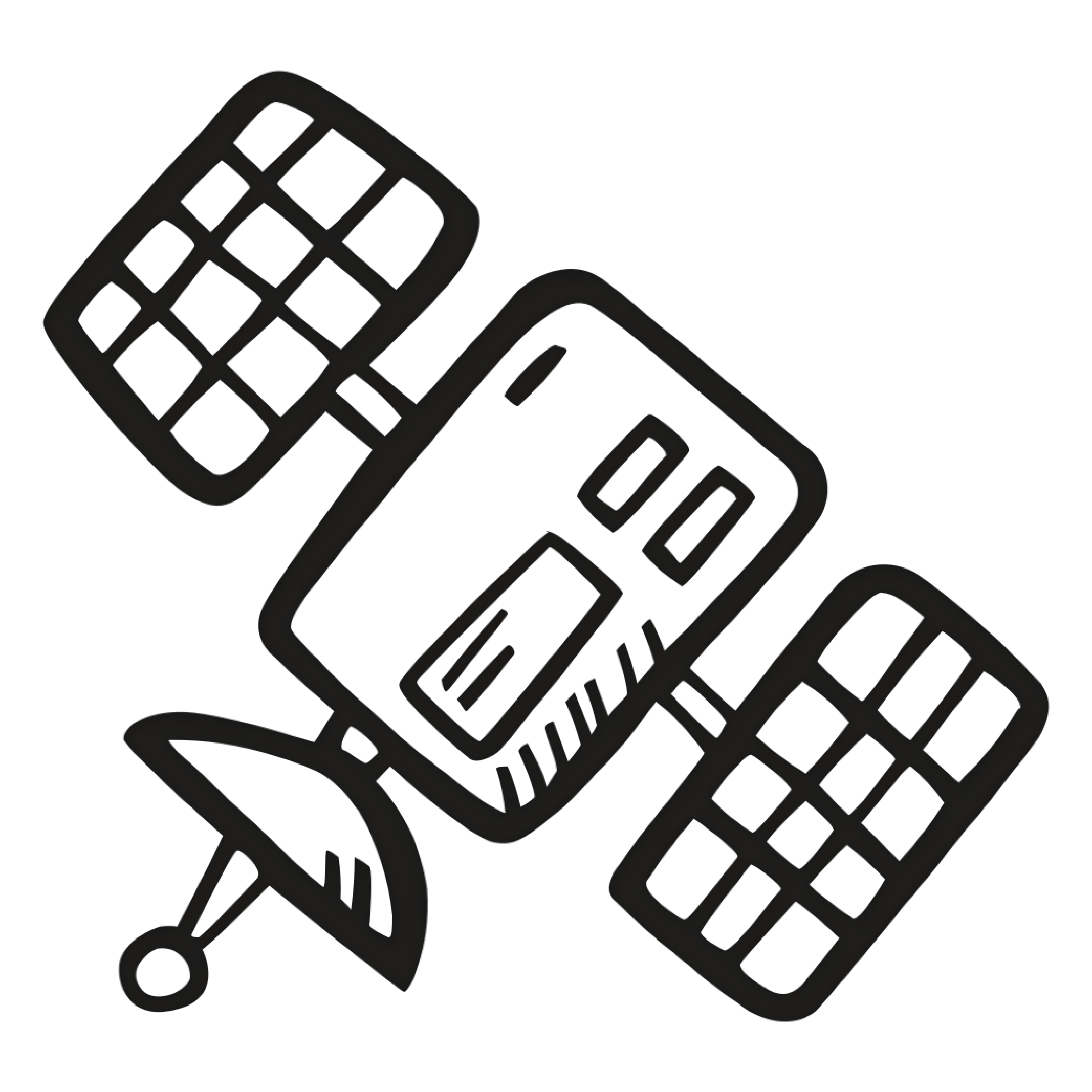}};
        \node[right] at (4.7,6) {Satellite};
        
        % NTN-UE
        \node (ntnue) at (1,3) {\includegraphics[width=1.5cm]{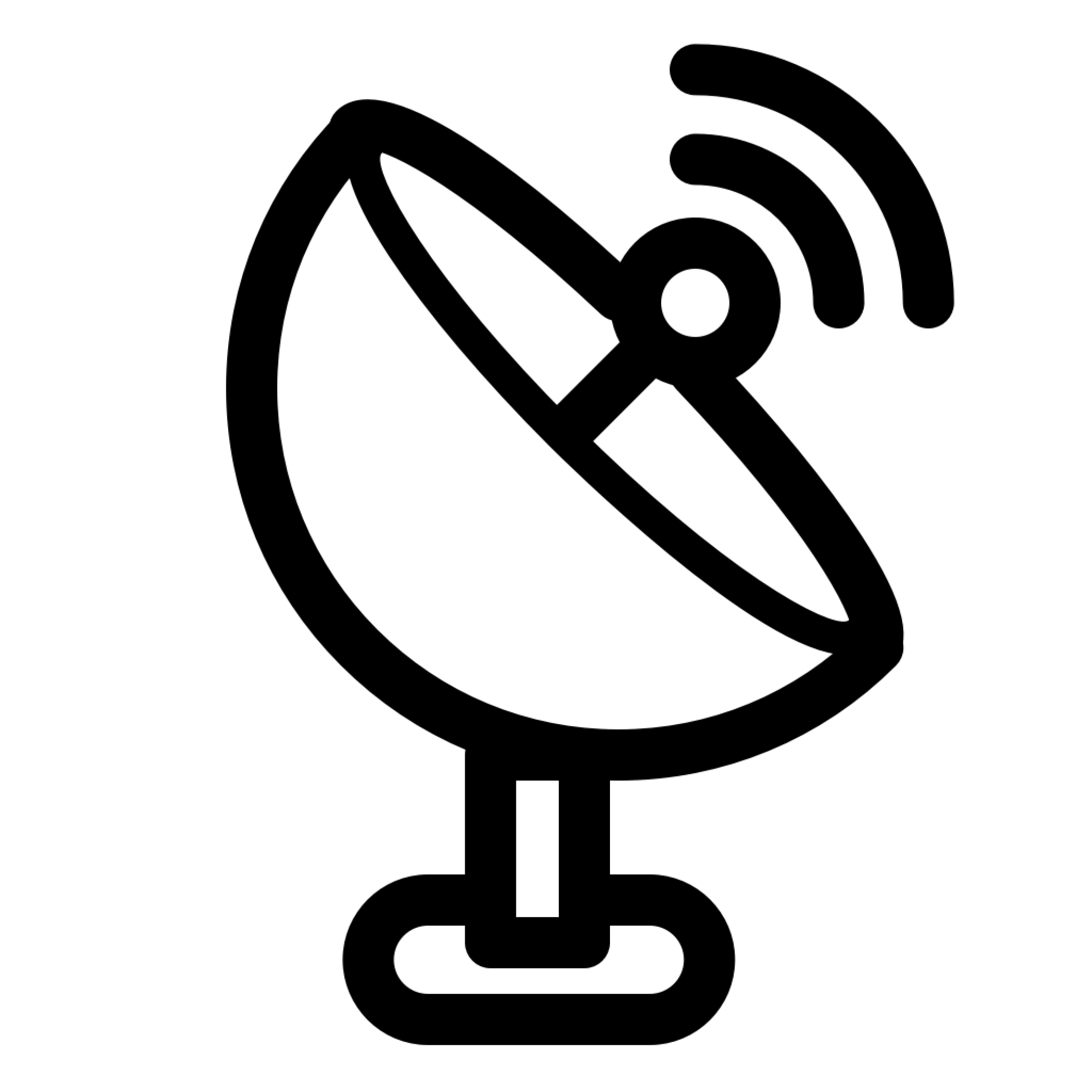}};
        \node[left] at (0.5,3) {NTN-UE};
        
        % Terrestrial gNB
        \node (gnb) at (7,3) {\includegraphics[width=1.5cm]{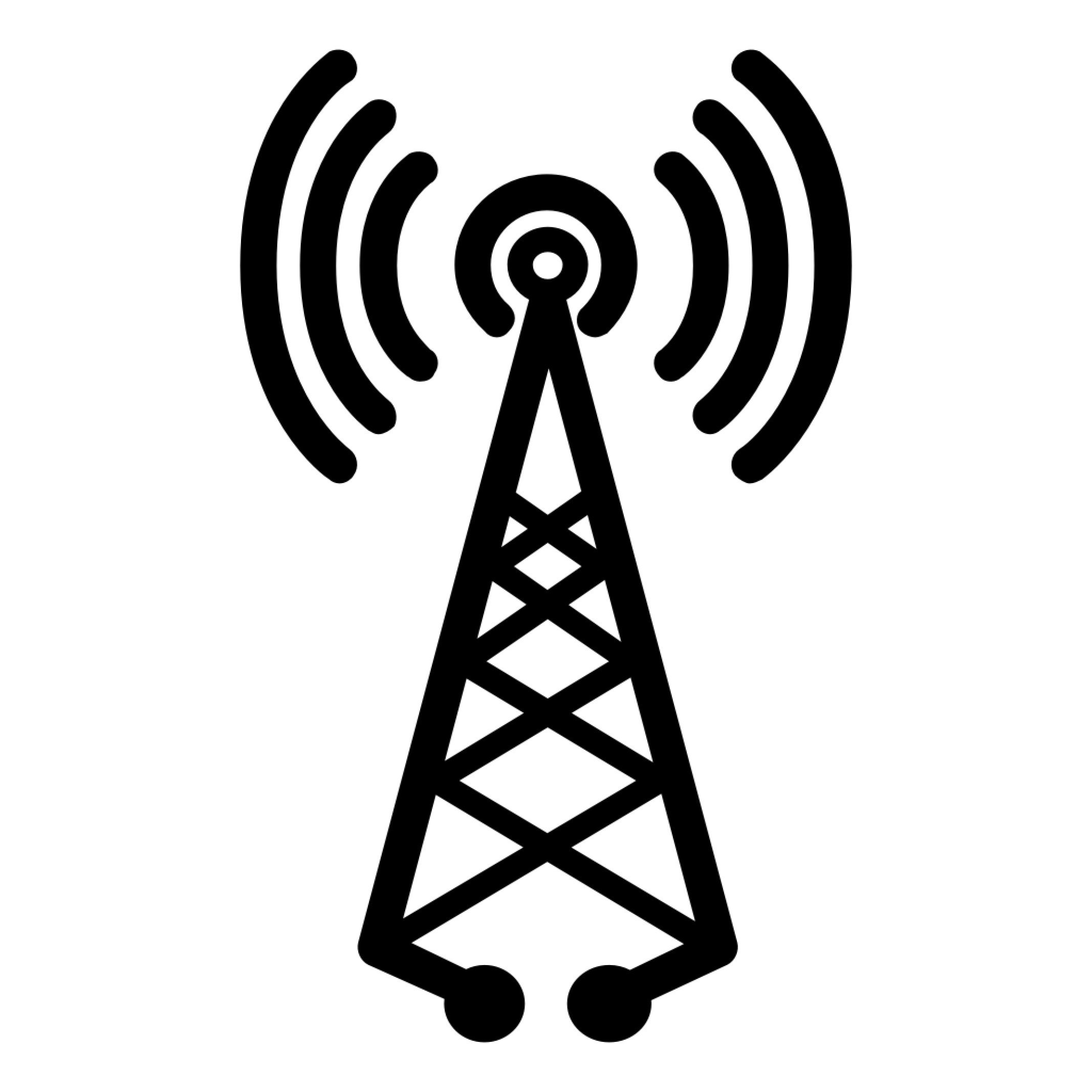}};
        \node[right] at (7.3,3) {TN-BS};
        
        % Terrestrial UE
        \node (terue) at (2,0) {\includegraphics[width=1cm]{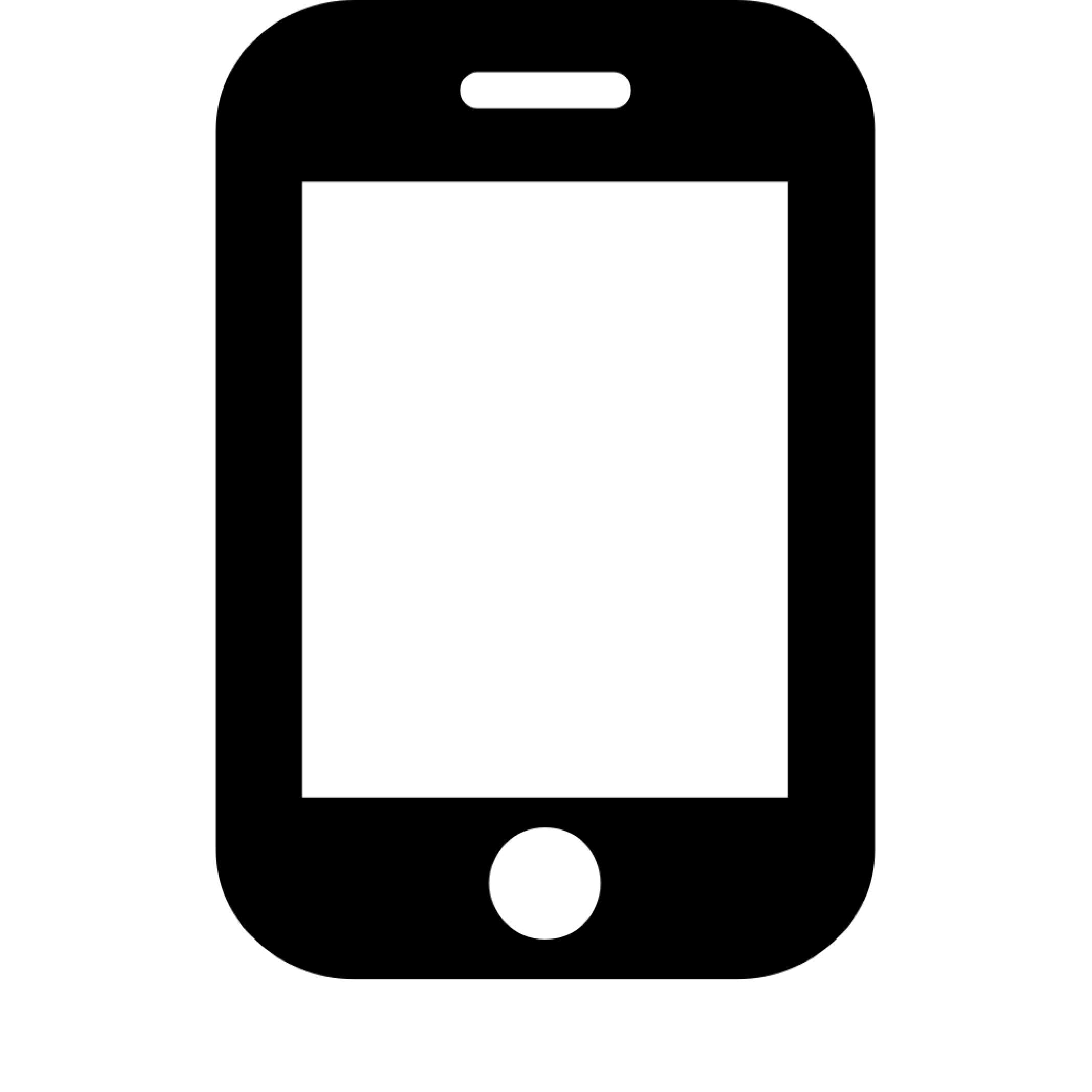}};
        \node[left] at (1.5,0) {TN-UE};
        
        % Separate arrows between NTN-UE and gNB with numbered labels
        \draw[->, thick, blue] (ntnue) to[bend left=5] node[midway, above] {1. Broadcast beacon} (gnb);
        \draw[->, thick, red] (gnb) to[bend left=5] node[midway, below] {4. Nulling} (ntnue);
        
        % Other arrows
        \draw[->, thick, ForestGreen] (satellite) -- (ntnue) node[midway, above] {};
        \draw[->, thick, ForestGreen] (gnb) -- (terue) node[midway, above, sloped] {4. Beamform};
        
        % Add text annotations near the relevant components
        \node[right, align=left] at (6.5,1.8) {2. Detect victim\\3. Estimate channel};
    \end{tikzpicture}
    \caption{Proposed interference mitigation framework for non-terrestrial and terrestrial co-existence networks: (1) NTN-UE transmits beacon signal, (2) TN-BS detects potential interference victim, (3) TN-BS estimates channel between TN-BS and NTN-UE, and (4) TN-BS implements spatial filtering with simultaneous null steering toward NTN-UE and optimized beamforming to TN-UE, maximizing spectrum sharing efficiency.}
    \label{fig:interference_mitigation}
    \vspace{-5pt}
\end{figure}

Previous research has extensively addressed interference mitigation, particularly focusing on the terrestrial-to-satellite uplink scenario \cite{10622881}. Explored techniques include interference nulling from TN base stations (TN-BSs) to satellites \cite{10622881}, 3D beamforming with feed array reflectors for spectral coexistence \cite{sharma20153d}, using transmit beamforming at the TN-BS to protect satellite terminals by leveraging their fixed pointing directions\cite{sharma2013transmit}, and secure beamforming for 5G mmWave systems coexisting with satellite networks \cite{lin2018robust}. Other efforts introduced theoretical interference models \cite{lim2023interference} and multicast beamforming strategies for cloud-based terrestrial-satellite architectures \cite{zhang2019multicast}. More recently, interference management in coexisting non-terrestrial networks (NTNs), such as dense low Earth orbit (LEO) constellations, has been studied. These include in-band coexistence via strategic satellite selection by secondary LEO systems to protect incumbents \cite{kim2024feasibility}, optimization and learning-based selection frameworks to maximize secondary system capacity under interference constraints \cite{kim2025satellite}.While these studies address uplink and inter-satellite interference, recent work has also examined the downlink case, notably proposing reverse spectrum pairing \cite{lee2023feasibility, lee2023interference}, in which TN uplink overlaps with NTN downlink on time. This paper focuses on the distinct challenge of satellite downlink interference, where TN-BS
transmissions interfere with NTN-UEs.

Satellite downlink presents fundamentally different challenges compared to uplink interference scenarios. Potential victim NTN-UEs remain unknown to TN-BS operators, with locations typically falling within the TN-BS antenna downtilt regions. NTN-UE channel information remains equally unknown, significantly complicating interference management. 
In contrast, satellite uplink interference management benefits from deterministic satellite positions available through ephemeris data. Satellites maintain relatively constant line-of-sight (LOS) paths, enabling easier detection and interference mitigation compared to the dynamic, unknown NTN-UE distribution in downlink scenarios.
Interference mitigation in the satellite downlink thus requires detecting potential victims and estimating the channels to those victims before interference nulling. 
% transmissions the channels
% Current approaches lack comprehensive frameworks addressing these dynamic interference scenarios and simultaneous TN-NTN requirements.

To address this challenge, our work introduces a joint beamforming optimization framework for TN-to-NTN interference mitigation, focusing specifically on TN-BS interference to NTN-UEs. The framework estimates interference channels to multiple NTN-UEs without precise location knowledge, enabling spatial nulling while preserving TN signal quality.

Key contributions include:

\begin{itemize}
    \item \emph{Detection and Nulling protocol:} We consider a four-stage protocol shown in Fig.~\ref{fig:interference_mitigation} where the potential NTN-UEs broadcast beacon or preamble signals.  From the preamble signals, the TN-gNB can detect the victim NTN-UEs and estimate the downlink channel.  After detecting and estimating the channels to the victims, the TN-gNB can beamform to the desired users with nulls in the directions to the victims.      
    \item \emph{Path loss effect on channel estimation:} Perfect channel knowledge enables complete signal nulling to victims. For imperfect estimation, path loss creates competing effects: increasing path loss improves victim detectability and channel estimation quality, while simultaneously amplifying residual interference impact. Analysis reveals critical detection-interference tradeoffs in single-user scenarios.
    
    \item \emph{Case Study:} We validate the performance in a rural area near Boulder, Colorado with realistic ray tracing simulations capturing complex propagation phenomena in rural macrocell (RMa) environments. Results confirm interference reduction effectiveness across varying antenna configurations and victim positions.
    
    \item \emph{Massive MIMO necessity:} Analysis demonstrates minimum antenna requirements (64+ elements) for effective null steering toward NTN-UEs while maintaining adequate TN-UE signal quality. Increasing antenna counts significantly enhances coexistence performance. 
\end{itemize}

%% file: problem.tex
\section{Problem Formulation}

\paragraph*{Network Model}    
We consider the scenario of a terrestrial base station TX wishes to transmit a downlink signal to a designated TN-UE, $\RX{0}$. Additionally, the network scenario includes $N$ victim receivers, labeled $\RX{1}$ through $\RX{N}$, which can be either Non-Terrestrial Network User Equipments (NTN-UE) or Non-Terrestrial Network Base Stations (NTN-BS). These victim receivers may reside on the ground or be airborne and suffering sidelobe interference. The base station is equipped with $N_\mathrm{tx}$ transmit antennas, while each receiver has a single antenna. The vectors $\bsym{h}_i\in \mathbb{C}^{N_\mathrm{tx} \times 1}$, where $i = 0, 1, \ldots, N$ denote the narrowband wireless channels between the base station and each receiver.

The transmission employs an $N_\mathrm{tx}$-dimensional beamforming vector $\bsym{w}$. Let $\mc{E}_{\subsf D}$ denote the downlink transmitted energy per symbol and $N_0$ represent the noise energy at each receiver, which is assumed uniform across all receivers. Given the selected beamforming vector $\bsym{w}$, the signal-to-noise ratio (SNR) at the intended receiver $\RX{0}$ is expressed as:
\begin{equation}\label{eq:snr}
    \SNR_0 = \frac{\mc{E}_{\subsf D}|\bsym{w}^*\bsym{h}_0|^2}{N_0}.
\end{equation}

Similarly, the interference-to-noise ratio (INR) experienced by each victim receiver $\RX{i}$, for $i = 1,\ldots,N$, is:
\begin{equation} \label{eq:inr}
    \INR_i = \frac{\mc{E}_{\subsf D}|\bsym{w}^*\bsym{h}_i|^2}{N_0}, \quad i=1, \ldots, N. 
\end{equation}

\section{Proposed Detection and Nulling Procedure}
The key challenge we address is that
the presence of the victim receivers
and the channels to the victims are not known.
We consider a simple four-part procedure, as shown in Fig.~\ref{fig:interference_mitigation}: (1) preamble transmission; (2) detection of the victims;
(3) estimation of the channel vectors to the victims; and
(4) interference nulling to the detected victims using the estimated channels.

\medskip
\paragraph*{Uplink preamble transmission}
To identify potential victims,
each victim RX selects one of
a finite set of known preamble sequences.
The sequence is broadcast so that the
transmitters can detect the presence
of the victims.
The TX performs a match filter
search for the preambles similar to what would be used in random access.
We assume the uplink channel is identical
to the downlink channel. 
Hence, on the correct delay and preamble hypothesis,
the receiver will receive a match filter output of the form:
\begin{equation} \label{eq:rpreamble}
    \bsym{r}_i = \bsym{h}_i a_i + \bsym{w}_i, \quad 
    \bsym{w}_i \sim {\mathcal CN}(0,N_0\bsym{I}), \quad i=1, \ldots, N
\end{equation}
where $a_i$ is the transmitted symbol
on the preamble. 
We let $\mc{E}_{\subsf U} = |a_i|^2$ denote the total energy
on the uplink preamble sequence.
Note that $\mc{E}_{\subsf  U}/\mc{E}_{\subsf D}$
may be large, since $\mc{E}_{\subsf D}$ represents
the energy on a typical sample,
while $\mc{E}_{\subsf U}$ is the energy on
a reference signal of many samples.
We also let:
\begin{equation} \label{eq:gammri}
    \gamma^{\subsf U}_i := \frac{\mc{E}_{\subsf U}G_i}{N_0}, \quad  G_i := \frac{1}{N_{\subsf tx}}\|\bsym{h}_i\|^2,
\end{equation}
so that $G_i$ is the average path gain per antenna and $\gamma^{\subsf U}_i$ is SNR during training for the $i$-th receiver per antenna.

\medskip
\paragraph*{Victim detection}
On each delay and preamble hypothesis, the base station
uses a simple energy threshold detector to determine if a potential victim exists.
Specifically, it assumes there is a victim at the delay when
\begin{equation}\label{eq:threshold}
    \|\bsym{r}_i\|^2 \geq t N_0,
\end{equation}
for some threshold $t$.  To select the threshold $t$, note that, when there is no signal, $\|\bsym{r}_i\|^2$ is
a scaled chi-squared random variable. 
 Hence, the threshold can be selected for directly for a given false alarm probability, $P_{\subsf FA}$.  As discussed in \cite{barati2016initial},
 the $P_{\subsf FA}$ value should be sufficient for the number of preamble, delay hypotheses, and carrier frequency hypotheses.

\medskip
\paragraph*{Channel estimation}
Having detected the potential victim receivers, the base station forms
of the channel to the receiver via equalizing the transmitted symbol:
\begin{equation}  
    \bsym{\widehat{h}}_i = \frac{a_i^*}{|a_i|^2}\bsym{r}_i.
\end{equation}
Observe that if the channel estimate is performed on a valid
preamble and delay hypothesis so that \eqref{eq:rpreamble} holds,
the channel estimate will be:
\begin{equation}       \label{eq:chanest}
    \bsym{\widehat{h}}_i = \frac{a_i^*}{|a_i|^2}\bsym{r}_i = \bsym{h}_i + 
    \bsym{v}_i,
\end{equation}
where $\bsym{v}_i$ is the channel estimation error,
\begin{equation} \label{eq:chanestvar}
     \bsym{v}_i \sim {\mathcal CN}\left(0, \frac{N_0}{\mc{E}_{\subsf U}}\bsym{I} \right),
\end{equation}

\medskip
\paragraph*{Interference Nulling}
Once the channels are estimated we then select the beamforming vector via
a regularized nulling:
\begin{equation} \label{eq:intnull}
    \wh{\bsym{w}} = \arg \max_{\|\bsym{w}\|=1} \left[ 
    |\bsym{w}^* \tilde{\bsym{h}}_0 |^2 - \lambda \sum_{i=1}^K 
     |\bsym{w}^* \wh{\bsym{h}}_i |^2\right],
\end{equation}
where $K$ is the number of detected victim receivers, and $\lambda \geq 0$ denotes a regularization parameter balances maximizing the desired signal power to TN-UE against suppressing interference toward NTN-UEs,
and $\tilde{\bsym{h}}_0 = \bsym{h}_0 /\|\bsym{h}_0\|$ is the normalized channel to the desired TN-UE.
To solve \eqref{eq:intnull}, $\wh{\bm{w}}_i$ is the principal eigenvector of the following Hermitian matrix:
\begin{equation} \label{solve bf}
    \bsym{Q} =  \tilde{\bsym{h}}_0 \tilde{\bsym{h}}_0^* 
    -\lambda \sum_{i=1}^K \wh{\bsym{h}}_i \wh{\bsym{h}}_i^*
\end{equation}
Note that a key appealing feature of
\eqref{eq:intnull} is that victims
that are further away (i.e., $\|\wh{\bsym{h}}_i\|$ is smaller) will naturally have smaller influence
in the nulling.

%% file: results.tex
\begin{figure}[!t]
    \centering
    \includegraphics[width=0.45\textwidth]{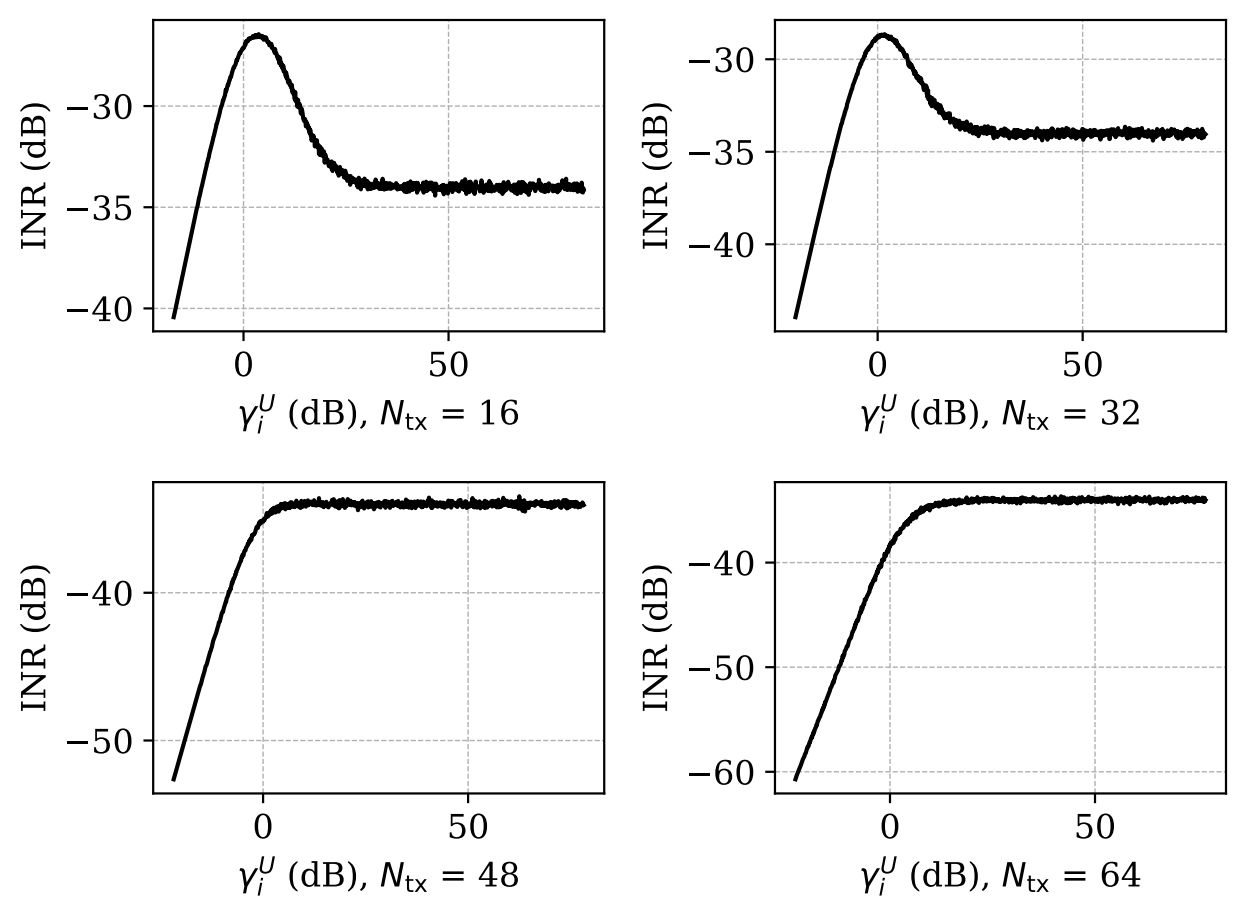}
    \caption{INR after interference nulling versus uplink detection receiver per antenna SNR $\gamma^{\subsf U}_i$ for different numbers of transmit antennas $N_{\mathrm{tx}}$.}
    \label{fig:single_victim_inr}
    \vspace{-10pt}
\end{figure}

\section{Practical Study and Performance Analysis}
\label{sec:performance_analysis}

\subsection{Single Victim Analysis}

Before considering a complex multi-user scenario, it is useful to first consider a single victim scenario to gain insights into the role of the path loss on the INR. We consider a single MIMO array with $N_{\subsf tx}$ antennas; the desired user is at boresight ($\theta = 0^\circ$ azimuth from the antenna), and a single victim user is at $\theta=15^\circ$. We assume the victim is detected and the base station attempts to perform the nulling \eqref{eq:intnull} with $\lambda \rightarrow \infty$ so the signal to the victim is completely nulled if there was no channel estimation error.

Fig.~\ref{fig:single_victim_inr} shows the INR as a function of the 
uplink SNR $\gamma_i^{\subsf U}$
with the uplink-to-downlink energy fixed at $\mc{E}_D/\mc{E}_U \approx $\,
\SI{-33}{dB} (the same values we use below).  We see an interesting effect:
initially, as $\gamma_i^{\subsf U}$
increases, the INR increases since the
victim is getting closer to the base station.  But, as it very close,
the INR saturates.  This saturation occurs since the channel estimation 
to the victim improves which compensates for the lower path loss. 
Also, when there are a smaller number
of antennas, there is a peak in the
INR that is caused since the base station cannot separate the victim and desired user well.

\subsection{Multi-Victim Scenario Evaluation}

\begin{figure}[!t]
    \centering
    \includegraphics[width=0.37\textwidth]{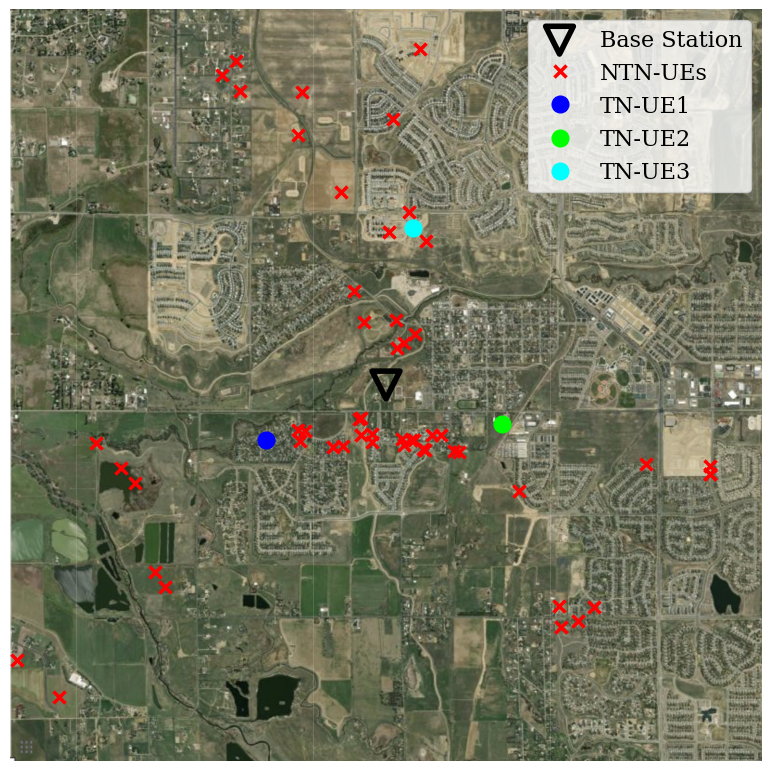}
    \caption{Simulation environment: \SI{6}{\kilo\meter} $\times$ \SI{6}{\kilo\meter} rural area near Boulder, Colorado.}
    \label{fig:map}
    % \vspace{-15pt}
\end{figure}

We next consider a more realistic 
multi-user scenario where a \SI{6}{\kilo\meter} $\times$ \SI{6}{\kilo\meter} rural area near Boulder, Colorado, as shown in Figure~\ref{fig:map}, serves as the simulation environment. A three-sector terrestrial base station (BS) is centrally positioned outdoors at an altitude of \SI{40}{\meter}. TN-UEs are randomly distributed within a range of \SI{35}{\meter}–\SI{3000}{\meter} from the BS, placed \SI{1.6}{\meter} above ground level, with 80\% located indoors. NTN-UEs are randomly deployed throughout the scene, with 80\% positioned \SI{1}{\meter} above rooftops and the remaining 20\% at \SI{1.6}{\meter} above ground level, outdoors. The channel is simulated using the open-source Sionna ray-tracing platform~\cite{sionna}.

All NTN-UEs are assumed to be aligned with a LEO satellite at a \SI{500}{\kilo\meter} orbit, with azimuth and elevation angles uniformly randomly distributed over $[0^\circ, 360^\circ]$ and $[25^\circ, 90^\circ]$, respectively, corresponding to BS. At each time instant, each sector randomly selects a TN-UE associated with that sector based on signal strength, assuming perfect channel state information (CSI) between the BS and TN-UEs.

% \begin{table}[htbp]
% \centering
% \caption{\textbf{Simulation Parameters.}}
% \label{tab:simulation_parameters}
% \renewcommand{\arraystretch}{1.3}
% \large
% \resizebox{0.48\textwidth}{!}{
% \begin{tabular}{|l|c|}
% \hline
% \textbf{Radio Parameters} & \textbf{Values} \\ \hline
% Carrier Frequency and Bandwidth & \SI{10}{\giga\hertz} and \SI{100}{\mega\hertz} \\ \hline
% Noise Power Spectral Density & \SI{-174}{dBm\per\hertz} \\ \hline
% Transmit Power & \SI{30}{dBm} (BS, VSAT)\\
% &\SI{23}{dBm} (Handheld) \\ \hline
% Noise Figure & \SI{3}{dB} (BS), \SI{2}{dB} (VSAT)\\
% &\SI{7}{dB} (Handheld) \\ \hline
% Maximum Number of Multipath & 3 (Reflection and Penetration) \\ \hline
% \multicolumn{2}{|l|}{\textbf{Antenna Type and Configuration}} \\ \hline
% Base Station (BS) & $4\times16$ URA, Directional \\ \hline
% Handheld (TN-UE) & (1, 1, 2) Omni-Directional~\cite{3gpp.38.821} \\ \hline
% VSAT (NTN-UE) & \SI{60}{\centi\meter} diameter, Directional \\ \hline
% Antenna Heights & \SI{40}{\meter} (BS), \SI{1.6}{\meter} (ground), \SI{1}{\meter} (rooftop) \\ \hline
% \multicolumn{2}{|l|}{\textbf{Scene Configuration}} \\ \hline
% Radio Material & Wall: Brick, Roof: Plywood \\
% &Ground: Medium Dry \\ \hline
% Satellite Angle & Azimuth: [0°, 360°], Elevation: [25°, 90°] \\ \hline
% Satellite Orbit & \SI{500}{\kilo\meter} (LEO) \\ \hline
% \end{tabular}
% }
% \end{table}

\begin{table}[!t]
\centering
\caption{\textbf{Simulation Parameters.}}
\label{tab:simulation_parameters}
\renewcommand{\arraystretch}{1.3}
\large
\resizebox{0.48\textwidth}{!}{
\begin{tabular}{|l|c|}
\hline
\textbf{Radio Parameters} & \textbf{Values} \\ \hline
Carrier Frequency and Bandwidth & \SI{10}{\giga\hertz} and \SI{100}{\mega\hertz} \\ \hline
Noise Power Spectral Density & \SI{-174}{dBm\per\hertz} \\ \hline
Transmit Power & \SI{30}{dBm} (TN-BS, VSAT), \SI{23}{dBm} (Handheld)\\ \hline
Noise Figure & \SI{3}{\dB} (TN-BS), \SI{2}{\dB} (VSAT), \SI{7}{\dB} (Handheld) \\ \hline
Maximum Number of Multipath & 3 (Reflection and Penetration) \\ \hline
Preamble Duration $T_{\subsf{pre}}$ & \SI{20}{\micro\second} \\ \hline
Probability of False Alarm $P_{\subsf{FA}}$ & $10^{-8}$ \\ \hline

\multicolumn{2}{|l|}{\textbf{Antenna Type and Configuration}} \\ \hline
Base Station (BS) & $4 \times 16$ URA, Directional~\cite{3gpp.38.901}, 3 sectors \\ \hline
Handheld (TN-UE) & (1, 1, 2) Omni-directional~\cite{3gpp.38.821} \\ \hline
VSAT (NTN-UE) & \SI{60}{\centi\meter} diameter, Directional \\ \hline
Antenna Heights & \SI{40}{\meter} (TN-BS), \SI{1.6}{\meter} (ground), \SI{1}{\meter} (rooftop) \\ \hline

\multicolumn{2}{|l|}{\textbf{Scene Configuration}} \\ \hline
Geographical Location  & \SI{6}{\kilo\meter} $\times$ \SI{6}{\kilo\meter} rural area \\ \hline
Radio Material & Wall: Brick, Roof: Plywood, Ground: Medium Dry \\ \hline
Satellite Angle & Azimuth: [0°, 360°], Elevation: [25°, 90°] \\ \hline
Satellite Orbit & \SI{500}{\kilo\meter} (LEO) \\ \hline
NTN-UE Deployment & 80 rooftop, 20 outdoor\\ \hline
TN-UE Deployment & 240 indoor, 60 outdoor \\ \hline
TN-BS Deployment & Centrally located, \SI{10}{\degree} downtilt \\ \hline
\end{tabular}}
% \vspace{-5pt}
\end{table}

\begin{figure}[!t]
    \centering
    \includegraphics[width=0.37\textwidth]{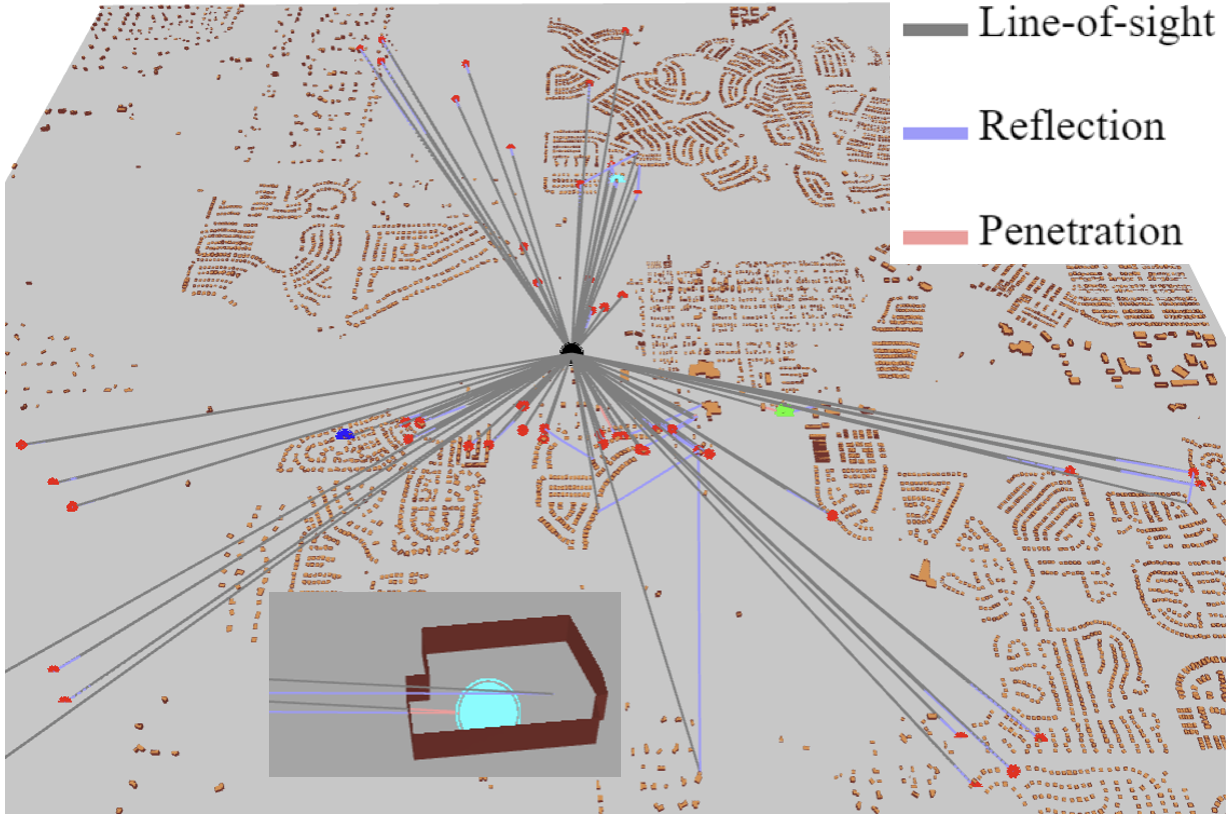}
    \caption{Ray tracing simulation in Sionna-RT showing line-of-sight (LOS), reflection, and penetration paths.}
    \label{fig:ray_map}
    \vspace{-15pt}
\end{figure}
\subsection{Terrestrial Downlink Interference Nulling}

\begin{figure}[!t]
  \centering
  \begin{subfigure}[b]{0.8\linewidth}
    \centering
    \includegraphics[width=\linewidth]{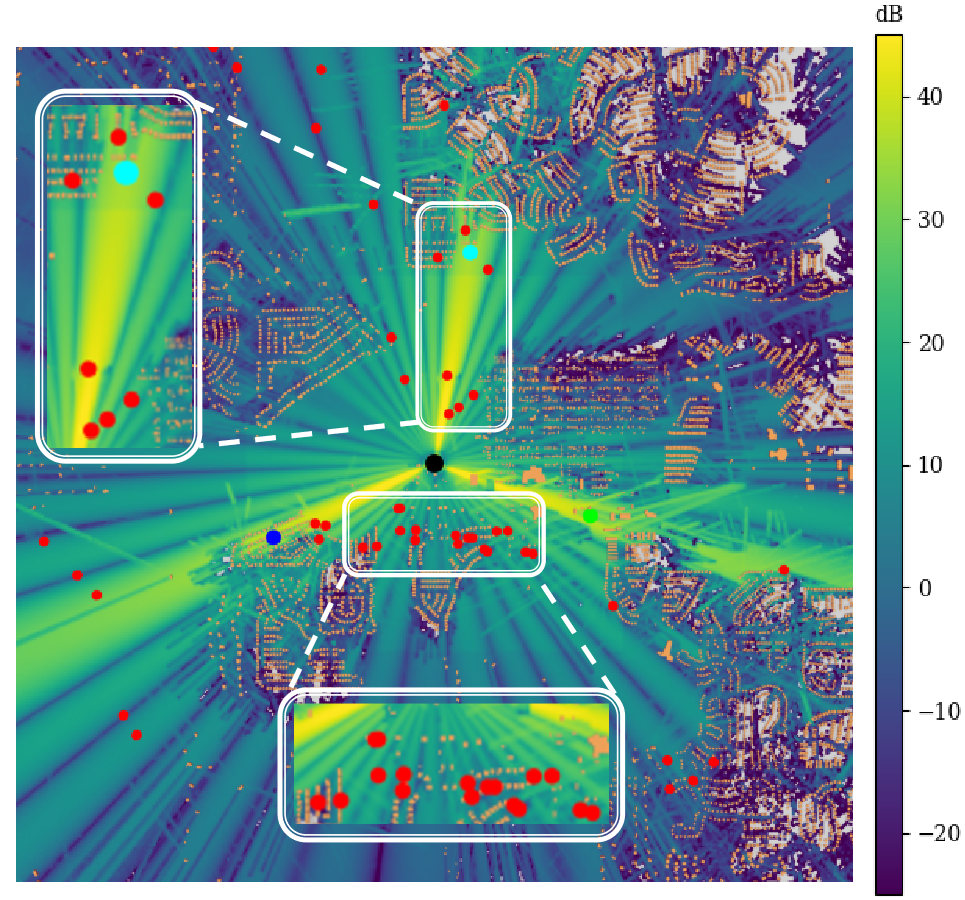} 
    \caption{}
    \label{fig:radio_map_a}
  \end{subfigure}

  \vspace{-0.1em}   

  \begin{subfigure}[b]{0.8\linewidth}
    \centering
    \includegraphics[width=\linewidth]{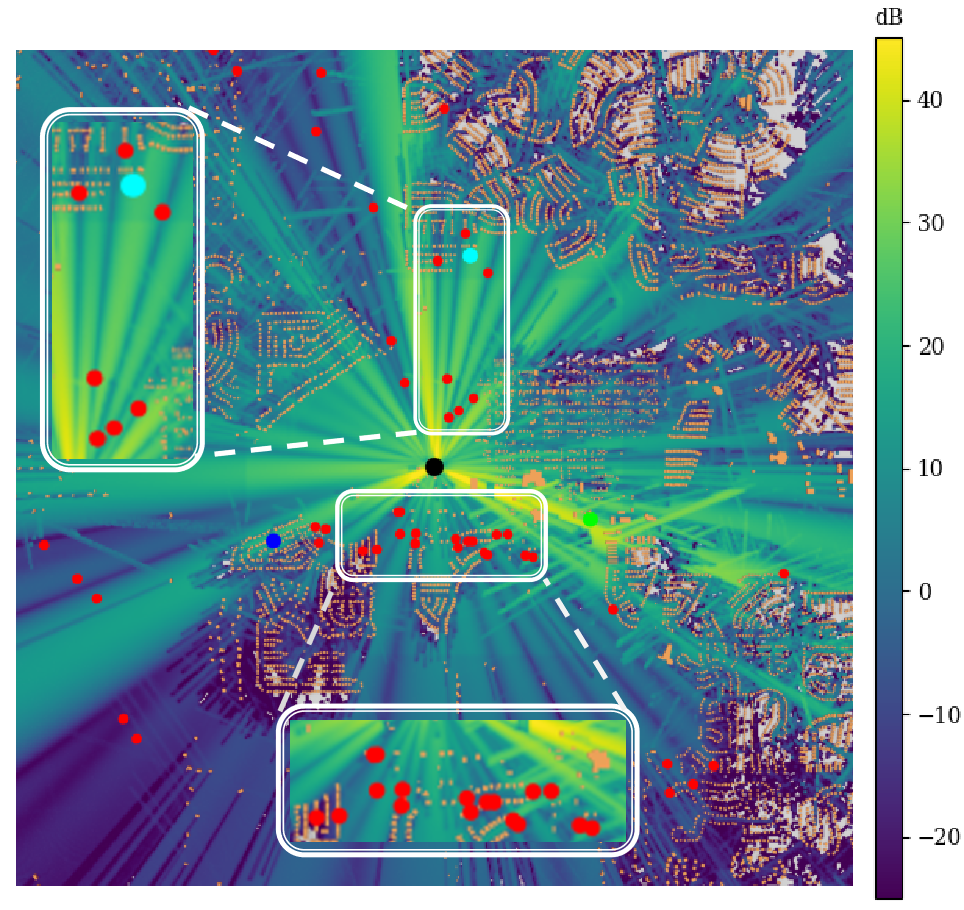}
    \caption{}
    \label{fig:radio_map_b}
  \end{subfigure}
  \caption{Simulated INR radio maps for the rural scenario. (a) shows the result before nulling, and (b) shows the result after nulling.}
  \label{fig:radio_maps}
  \vspace{-10pt}
\end{figure}

An example of the interference nulling
for one instance is shown in Fig.~\ref{fig:radio_maps}.
Fig.~\ref{fig:radio_map_a} shows the
coverage map with no interference nulling.  The signal to the desired user (blue is maximized), but there is significant interference at the victims (red).  In contrast,
when interference nulling \eqref{eq:intnull} is applied with $\lambda=10^{11}$, nulls on the victims, particularly those close to the base station, are formed. 

% \paragraph*{Baseline}
% As the baseline, we evaluate the interference experienced by NTN-UEs without applying interference nulling. In this case, the BS performs transmit beamforming using singular value decomposition (SVD) by solving (\ref{solve bf}) with $\lambda=0$ to its selected TN-UEs, as shown in Fig.\ref{fig:radio_map_a}, three strong beams towards their picked TN-UEs while NTN-UEs in the beam area.

% \paragraph*{INR nulling to NTN-UEs}
% One victim NTN-UE may suffered from and detected by more than one sector of BS, so the INR is calculated as (\ref{eq:inr}) but the sum of all interference signals over noise. Fig.\ref{fig:radio_map_b} shows the INR radio map after applying nulling method using (\ref{eq:intnull}), we successfully cause nulls to NTN-UEs around targeted TN-UEs and compress INR from around \si{20}{dB} to below \si{-10}{dB} like the victims in the lower left part from radio map.

\begin{figure}[!t]
    \centering
    \includegraphics[width=0.4\textwidth]{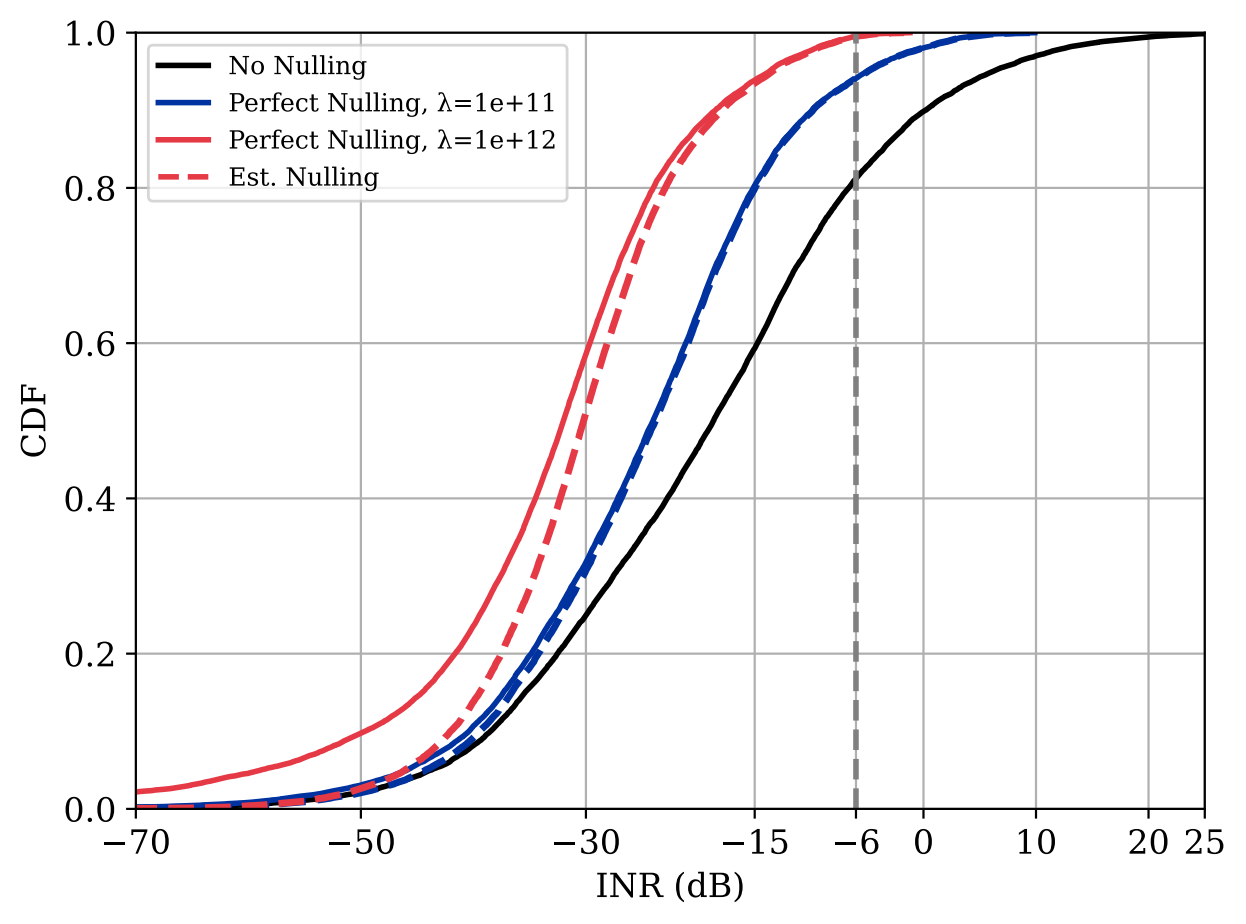}
    \caption{CDF of INR towards NTN-UEs with and without interference nulling. The black curve denotes the case without nulling. Colored curves correspond to different values of the regularization parameter~$\lambda$. Solid lines use perfect channel information, while dashed lines use estimated channels.}
    \label{fig:inr_cdf}
    \vspace{-10pt}
\end{figure}

Fig.~\ref{fig:inr_cdf} presents the cumulative distribution function (CDF) of the INR on the terrestrial downlink, evaluated through 100 Monte Carlo simulations. In each simulation, an average of 10 NTN-UEs is detected per sector. When interference nulling is applied, two regularization values are considered: $\lambda = 10^{11}$ and $\lambda = 10^{12}$. As observed from the no-nulling baseline (black curve), approximately 20\% of the detected NTN-UEs experience interference with an INR above \SI{-6}{dB}. With $\lambda = 10^{11}$ (blue curve), this percentage is reduced to around 8\%. For $\lambda = 10^{12}$ (red curve), the INR distribution shifts significantly to the left of \SI{-6}{dB}, indicating more effective interference suppression.

The dashed red curve corresponds to the estimated nulling case using the channel estimate $\bsym{\widehat{h}}_i$ from \eqref{eq:chanest}, obtained with a \SI{20}{ms} preamble time. In contrast, solid curves represent ideal performance assuming perfect channel knowledge $\bsym{h}_i$. For $\lambda = 10^{11}$, the estimated and perfect nulling curves align closely for INR values above \SI{-30}{dB}, with divergence appearing only at lower INR levels. In the case of $\lambda = 10^{12}$, performance degradation with estimated channels begins earlier (around \SI{-15}{dB}). For very low INR values, the estimated nulling performance converges toward the no-nulling case due to the noise floor limitation.

 \paragraph*{SNR degradation to TN-UEs}

\begin{figure}[!t]
    \centering
    \includegraphics[width=0.4\textwidth]{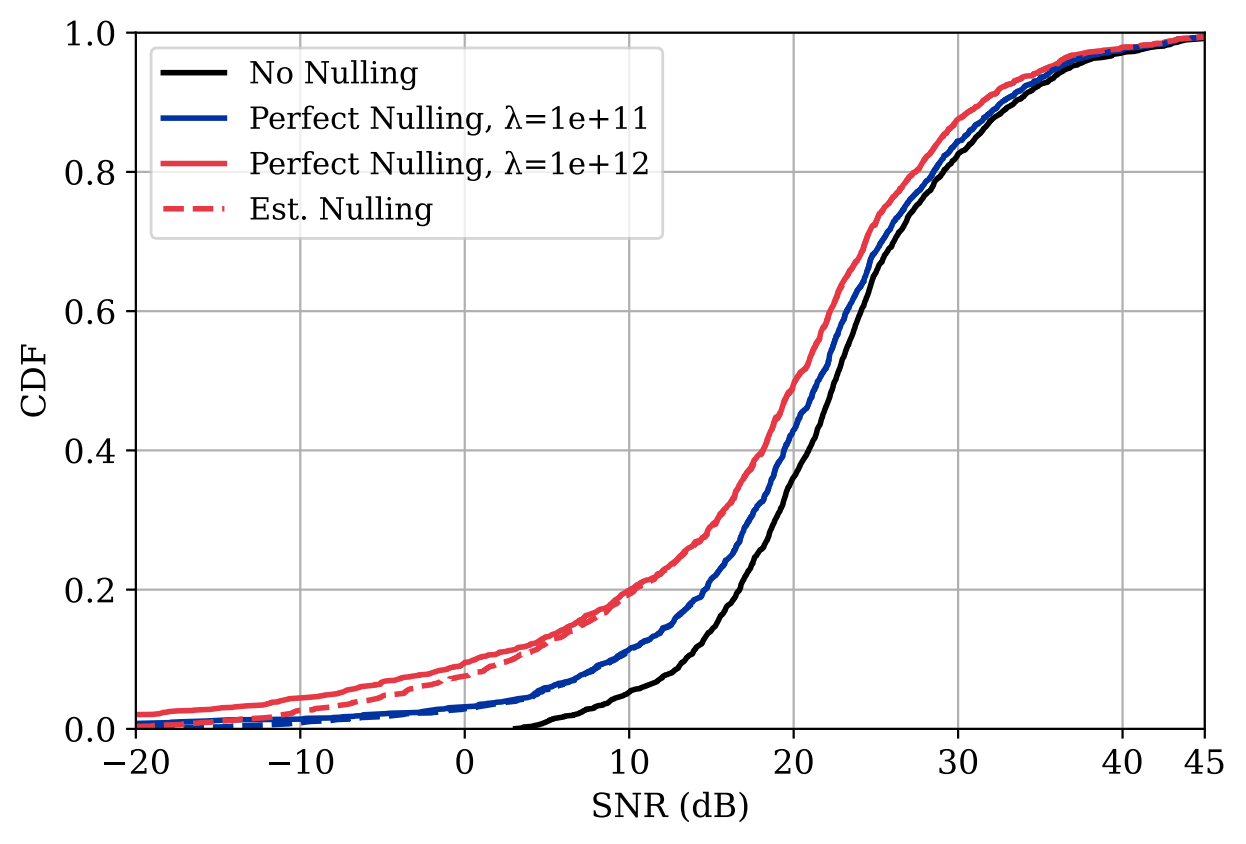}
    \caption{CDF of SNR towards TN-UEs under the same nulling configurations as in Fig.~\ref{fig:inr_cdf}.}
    \label{fig:snr_cdf}
    \vspace{-10pt}
\end{figure}

Fig.~\ref{fig:snr_cdf} shows the CDF of the SNR at the BS with and without interference nulling, where the nulling method from \eqref{eq:intnull} is applied to the SNR expression in \eqref{eq:snr}. As expected, applying nulling introduces some degradation to the SNR of the TN-UEs due to the projection of the beamforming vector onto the null space of the interference channel. Compared to the no-nulling case (black curve), the SNR distributions shift leftward under nulling schemes, indicating reduced signal strength at the TN-UEs.

For $\lambda = 10^{11}$ (blue curve), the degradation is moderate, and the SNR remains above \SI{0}{dB} for nearly all users. When increasing the regularization parameter to $\lambda = 10^{12}$ (red curve), the SNR loss becomes more pronounced, especially in the lower SNR range. The dashed red curve corresponds to the case with estimated channel $\bsym{\widehat{h}}_i$ using a \SI{20}{ms} preamble. Its performance closely follows that of perfect nulling for most of the SNR range, except in the very lower SNR region (below \SI{5}{dB}), where estimation errors lead to larger degradation.

Despite the observed loss, the nulling schemes still maintain acceptable SNR levels for the majority of TN-UEs while providing significant INR suppression, as shown in Fig.~\ref{fig:inr_cdf}. This highlights the trade-off between interference mitigation and signal quality in practical systems.

\subsection{Massive MIMO needs for Rising NTN Load}

To achieve effective interference suppression for an average of 10 detected NTN-UEs per sector, a relatively large regularization value of $\lambda = 10^{12}$ is required, as shown by the red curve in Fig.~\ref{fig:inr_cdf}. While this leads to some degradation in SNR performance, Fig.~\ref{fig:snr_cdf} shows that the impact remains within an acceptable range under such settings.

\begin{figure}[!t]
    \centering
    \includegraphics[width=0.4\textwidth]{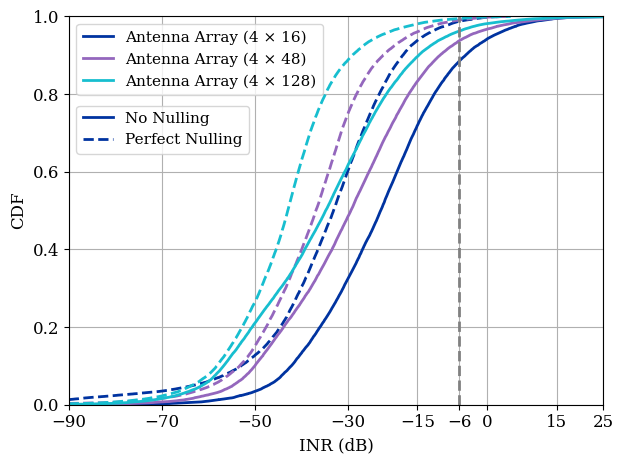}
    \caption{CDF of INR towards NTN-UEs for different antenna array size configurations, comparing performance with and without nulling under estimated and perfect channel information, ($\lambda=10^{11}$).}
    \label{fig:inr_cdf_antenna}
    % \vspace{-10pt}
\end{figure}

\begin{figure}[!t]
    \centering
    \includegraphics[width=0.4\textwidth]{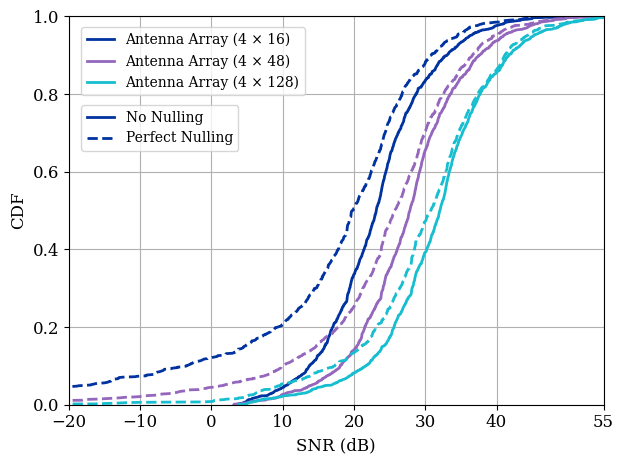}
    \caption{CDF of SNR towards TN-UEs for different antenna array size, corresponding to the same nulling configurations as in Fig.~\ref{fig:inr_cdf_antenna}.}
    \label{fig:snr_cdf_antenna}
    % \vspace{-15pt}
\end{figure}

As non-terrestrial user density continues to increase in future deployments—e.g., with 30 detected NTN-UEs per sector—the cumulative strength of the interference channel (the second term in \eqref{eq:intnull}) also increases. In such scenarios, a smaller regularization value such as $\lambda = 10^{11}$ may suffice to achieve the same interference suppression effect, as shown in the blue curve of Fig.~\ref{fig:inr_cdf_antenna}. This, however, comes at the cost of significantly greater SNR degradation, as evident from the corresponding blue curve in Fig.~\ref{fig:snr_cdf_antenna}. In these figures, solid lines represent the no-nulling baseline, while dashed and dash-dot lines correspond to interference nulling using estimated and perfect channel information $\bsym{\widehat{h}}_i$, $\bsym{h}_i$.

To address this growing trade-off between interference nulling and desired signal quality, large-scale (massive) MIMO systems offer a promising solution. By increasing the number of antenna elements at the BS, the system gains additional spatial degrees of freedom. This allows the BS to simultaneously steer beams toward desired TN-UEs while placing deeper nulls toward interfering NTN-UEs, thereby reducing the reliance on high regularization values and mitigating the associated SNR loss.

As shown in both Fig.~\ref{fig:inr_cdf_antenna} and Fig.~\ref{fig:snr_cdf_antenna}, moving from a $(4 \times 16)$ to a $(4 \times 128)$ antenna array significantly improves both INR suppression and SNR retention under nulling. In particular, the dashed curves shift leftward in Fig.~\ref{fig:inr_cdf_antenna} (stronger interference mitigation), while maintaining close proximity to the solid curves in Fig.~\ref{fig:snr_cdf_antenna} (limited SNR loss). These results clearly demonstrate that massive MIMO architectures are essential for maintaining downlink quality in high-density hybrid NTN–terrestrial environments, where the number of nulling constraints may grow substantially.

% \vspace{3\baselineskip}

% \begin{figure}[!t]
%     \centering
%     \includegraphics[width=0.45\textwidth]{results_pic/snr_deg_antenna.pdf}
%     \caption{CDF of SNR degradation for different antenna configurations under perfect channel estimation ($\lambda=10^{11}$).}
%     \label{fig:snr_deg_cdf}
% \end{figure}